# A compact inertial nano-positioner operating at cryogenic temperatures

Pritam Das[*], Sulagna Dutta[*], Krishna K.S[†]., John Jesudasan[‡] and Pratap Raychaudhuri[§]

*Tata Institute of Fundamental Research, Homi Bhabha Road, Mumbai 400005.*

Nano-positioning plays a very important role in applications such as scanning probe microscopy and optics. We report the development of a compact inertial nanopositioner along with fully computer interfaced electronics operating down to 2 K, and its use in our fully automated needle-anvil type Point Contact Andreev Reflection (PCAR) apparatus. We also present the fully automated operational procedures using LabVIEW interface with our home-built electronics. The point contact spectroscopy probe has been successfully used to perform PCAR measurements on elemental superconductors at low temperatures. The small footprint of our nano-positioner makes it ideally suited for incorporation in low temperature scanning probe microscopes and makes this design versatile for various research and industrial purposes.

[*] The authors contributed equally to this work
[†] Permanent address: Centre for Integrated Studies, Cochin University of Science and Technology, Kochi-682022, India.
[‡] E-mail: johnj@tifr.res.in
[§] E-mail: pratap@tifr.res.in

## I. Introduction

Precision positioning is an indispensable part of diverse fields from industrial applications[1,2,3], like in optics[4] and photonics[5], nanotechnology[6,7,8], metrology[9,10,11], bioengineering[12], semiconductor manufacturing[13], and aerospace engineering[14,15] in research and development purposes. These devices leverage the piezoelectric effect, where certain materials deform proportionally to the electric field applied to them and this property enables them to achieve exceptionally fine movements. It facilitates meticulous manipulation, measurement, and assembly at the nanoscale level with their compact size, sub-nanometer resolution[16] and low-power consumption, thereby propelling forward technological advancements and research endeavors. There are a variety of piezoelectric positioners that have been developed to achieve sub-nanometer precision[16]. Piezoelectric tube nano-positioners[17,18,19,20] and piezoelectric stack nano-positioners[21,22,23] offer sub-angstrom resolution, making them highly precise for nano-scale positioning tasks. However, this exceptional precision comes with a trade-off as they have a relatively small total translational motion range. When coupled with a coarse positioner, the movement range can be extended to several millimetres. This combination allows for both high precision and a larger translational movement, enabling precise positioning over a broader range.

The challenge is often to make low-footprint course positioner operating down to cryogenic temperatures[24,25] for various applications such as Scanning Probe Microscopy[26,27] and Optics[4]. Here, we describe the construction of our small footprint cryogenic nanopositioner which has a compact size, low making cost and easy assembly and efficiently operates in a range up to ~ 4 mm. We describe both the hardware of the inertial nanopositioner, which operates on the principle of slip-stick motion, and the fully automated electronics that control it, utilizing a microcontroller-based architecture for seamless computer integration. We tested the positioner by fabricating a point contact spectroscopy setup and measured the point contact spectra of Nb films by Cu tips down to ~ 2.5 K.

## II. Construction of the Inertial Nanopositioner

### A. Hardware

Our inertial nano-positioner has a very compact and simple design. We first describe the construction of Z nanopositioner for the movement in vertical direction. It has footprint of $15\ mm \times 15\ mm$ and a weight of 12.8 gram, thus can be easily integrated in different systems with various applications. The basic design has three principal parts. The active part is a piezoceramic stack (TA0505D024 from Thorlabs) which is glued to a moving stage on one side with Loctite-EA1C and the other side is glued to a triangular metallic monorail, acting as the guiding rod. This monorail slides between two blocks with matching groove as shown in Fig. 1. The frictional force between the two blocks and the triangular monorail can be adjusted by tightening the spring-loaded screws. Titanium is employed in the construction of this nano-positioner. A coating of finely crushed Molybdenum disulphide (MoS$_2$) is applied on the sliding surface of the monorail and the block to reduce friction. This ensures smooth motion of the monorail in the nanopositioner. The MoS$_2$ coating also acts as a dry lubricant to reduce the wear and tear on the side wall of the linear slide. We had also fabricated these parts with Aluminium. However, those nanopositioners tended to get stuck at low temperature probably due to the softness of Aluminium. Distance between the monorail and the moving stage can be varied by applying voltage to the piezo. The piezo stack expands when a positive voltage is applied between the two electrodes. In actual operation, sawtooth waveform is applied to the piezo by the power supply. First, the voltage is gradually increased, causing the piezo to expand slowly (~ 37 nm/V) at room temperature. This expansion moves up the movable stage, while the guiding rod adheres to the monorail holding block. Subsequently, the voltage drops very rapidly over a period of microseconds. Due to the rapid acceleration of the moveable stage in this step, the reaction force, M$a$, (where M is the mass of the moveable stage and $a$ is the acceleration) on the triangular monorail exceeds the static friction between the monorail and the holding blocks, and the monorail slides by a finite amount. This is illustrated in Fig. 2. Repeating these steps, we can move the stage to several mm distances in steps of tens of nanometers. For moving in the opposite direction, we simply must reverse the sawtooth, causing fast expansion and slow contraction of the piezo stack. The step-size during motion is controlled by the amplitude of the sawtooth voltage applied to it. In our nanopositioner, we observed that a minimum of 6V was needed to move the stage at room temperature. However, in actual operation we normally used between 10-60 V to move the stage over the entire range of 0-4 mm.

By using the same slip stick mechanism but a slightly modified design we have also made X nanopositioner that can be stacked on the Z-positioner using an L shaped attachment and shown in Fig. 1. Here the monorail is horizontal such that the stage translates in the horizontal direction. The X nanopositioner is also light weight ~ 12.7 grams. The stage can be moved to several mm distances in steps of tens of nanometer by the same principle as Z nanopositioner.

## B. Electronics of the Power supply

Generating sawtooth waveform with an amplitude ranging from 6 V to 60 V, is essential to drive the piezoelectric element of the positioner. We generate a sawtooth waveform with amplitude in the range 0-3.3V using the analog output from the Adafruit Feather M4 Express microcontroller. This signal is then amplified using a non-inverting amplifier. One of the key challenges is that the rise or fall time of the sawtooth waveform, should be very sharp (<10 microseconds) for the stage to move. During this step there is a rapid rate of change of voltage ~6 $V/\mu s$. Due to the high capacitance of the piezo stack ( $C \approx 1 \mu F$ ) this translates into a transient current, $I = C \times \frac{dV}{dt} \sim 6$ A. To accommodate such huge currents we use, the OPA541 power OP-amp for the non-inverting amplifier capable of operating up to 60V, 10A. Since we are only interested in generating positive voltage, we power the OP-amp using +60 V (+$V_{cc}$) and -8V (-$V_{cc}$). The power supply to the Op-amp is designed using two step-down transformers and rectifier circuits to give positive and negative voltages.

The feedback loop of the non-inverting amplifier consists of an 82 kΩ feedback resistor, whereas an input resistor of 4 kΩ is chosen to act as a gain limiter to ensure the gain remains at ≤ 20. In addition, a 5kΩ potentiometer is attached in series with the 4 kΩ resistor to reduce the gain as needed. We can control the amplitude of the sawtooth, either by adjusting this potentiometer or by digitally changing the amplitude of the waveform produced by the microcontroller. The circuit is shown in Fig. 3. A separate section of the circuit supplies 5V power to a digital voltmeter on the front panel, which displays the output voltage from the Op-amp. This output is directly fed to the piezoelectric element of the nanopositioner to move.

### C. Computer interface

To control the nanopositioner using a computer, we use the serial port of the Adafruit Feather M4 Express microcontroller. A single serial string sent from the computer to the microcontroller (9600 baud rate) contains information on the mode of operation (single step or continuous movement), amplitude and frequency of the sawtooth, and whether the positioner should move up or down. For the user interface we can use LabVIEW or Python. Details of the LabVIEW interface is given in the Supplementary Material.

The analog output of Adafruit Feather M4 Express microcontroller has 12 bits resolution to have a fine control over the output voltages. When a new data is detected at the serial input, the output of the amplifier is modified according to the latest string of commands. The structure of the string is explained in Appendix 1. Then it generates a continuous or single sawtooth waveform for forward or reverse movement with the desired amplitude and frequency. The timing between these steps is controlled by a delay which is calculated based on the user-provided frequency and amplitude values. This enables the waveform to be fine-tuned in real-time according to the input commands. By adjusting the delay, the code ensures that the output waveform can be dynamically modified, allowing for precise control over both the shape and timing of the signal. The waveforms for each mode are shown in Fig. 4. The Arduino IDE sketch for generating the waveforms is given in the supplementary material. This overall design allows for flexible and precise control of the waveform output, making it suitable for applications where custom signal generation is needed.

## III. Characterisation of the inertial nanopositioner

### A. Room temperature Characterisation

For optimal operation, the spring-loaded screws on the monorail holding block, must be carefully adjusted as shown in Fig. 1. If the screws are tightened too hard the vertical triangular monorail will not slide. On the other hand, if the screws are too loose the moving stage of the Z nanopositioner will come down under its own weight, especially with load. We found that the mass of our platform, 2.3 gms, provides sufficient inertial mass for the positioner to operate in unloaded condition. On the other hand, the maximum load that we can place on the platform is ~40 gms. To adjust the spring tension, we place 40 gms weight on the positioner platform at room temperature and adjust the spring tension by trial and error to ensure smooth forward and backward motion. We found that spring tension optimised in this way also ensures optimal motion at low temperature.

The comprehensive characterisation of the Z and X positioners at room temperature are shown in Fig. 5 and Table 2. For both nanopositioners, the extension was measured up to a maximum of 4 mm (with respect to the fully retracted position) after applying a given number of sawtooth pulses of different amplitudes and under different loads. The movement is linear as a function of the number of sawtooth pulses. However as expected, due to gravity, the Z-nanopositioner moved downward with larger movement per step than going up. This difference becomes large for larger amplitude of the sawtooth voltage and larger load on the positioner. In contrast, for the X positioner for both forward and reverse step size is same as expected.

### B. Low temperature characterisation

Below room temperature, the expansion coefficient of piezo elements drops rapidly. The rate of drop varies based on the piezo material; typically, below 4 K the expansion coefficient is 5 to 10 times smaller than that at room temperatures. Consequently, the step size of the nanopositioner gets smaller at low temperatures. For the piezo stacks used in our nanopositioners, we do not have the expansion coefficient data at cryogenic temperatures. Therefore, to test the performance of our nanopositioner at low temperatures, we incorporated the Z-positioner in a home-built Point Contact Spectroscopy insert[28,29,30,31] operating down to 2 K. In this technique, a fine metallic tip attached on the nanopositoner is brought towards a sample to eventually make a nanometer-sized ballistic contact between the tip and the sample. When the contact is in the ballistic limit the current voltage characteristics of such a contact carries spectroscopic information of the sample.

The schematic of the probe is shown in Fig. 6(a). In our setup a tip holding assembly (4.6 gm) is fixed on top the inertial Z nanopositioner. Two electrical leads are connected to the sample and two other electrical leads are connected to the tip and tip holder as shown in Fig. 6(b). This allows us 4-probe measurements of the resistance of the contact between the tip and the sample as long as the bulk resistance of tip wire is much smaller than the resistance of the contact. The nano positioner is taken forward towards the sample (placed upside down) in steps and the resistance between the sample and the tip is measured after each step. When the tip touches the sample surface the resistance drops abruptly showing that a contact is established. This probe goes inside an Oxford Instruments continuous flow [4]He cryostat. The distance between the tip with the z-positioner in fully retracted position is ~1 mm. To characterise the positioner, we compared the number of steps required to engage the tip from fully retracted position with sawtooth amplitude of 34 V at room temperature and at 2.5 K. From this ratio we inferred that the step size at 2.5 K is 5.2 times smaller than at room temperature.

### C. Point contact Andreev Reflection measurements

To demonstrate the performance of our walker in a real experiment, we performed Point Contact Andreev Reflection (PCAR) [32,33,34,35,36,37,38,39] measurements using a mechanically cut Cu tip and a superconducting Nb thin film grown through dc magnetron sputtering. Andreev reflection [40] is a process where an electron incident from the normal side of a normal metal/superconductor (N/S) interface with energy less than the superconducting energy gap, $\Delta$, gets reflected as a hole, whereas a Cooper pair propagates inside the superconductor. PCAR measurement is performed by establishing a ballistic contact (i.e. contact diameter smaller than the electronic mean free path), and measuring the conductance, $G(V) = \frac{dI}{dV}$ of the junction as a function of applied voltage, $V$, across the junction. For an ideal N/S interface with infinite transparency Andreev reflection process results in a doubling of the junction conductance for $V < \Delta/e$ compared to $V \gg \Delta/e$ as shown in Fig. 6(c). However, this is hardly ever realised in a real point contact where there is always a finite interfacial barrier resulting either from surface imperfections (like surface oxide) or the Fermi velocity mismatch between the normal metal and superconductor. This results in less than doubling of the low bias conductance as well as the appearance of two symmetric peaks at bias voltages close to $|V| \approx \Delta/e$ as shown in Fig. 6(d).

Fig. 7 shows a set of $G(V) \ vs. V$ PCAR spectra taken at 2.5 K. We first engage the tip on the sample with sawtooth amplitude of 30 V using the procedure described before. The black curve shows the spectra taken just after engaging the tip. Here, in addition to the regular Andreev reflection feature we also observe two sharp dips at higher voltages (shown with arrows). These dips are non-spectroscopic features that arise in larger contacts when the current across the point contact exceeds the critical current of the superconductor, implying that the contact is not purely in the ballistic limit[41,42,43]. Therefore, we subsequently retract the tip in very small steps using sawtooth amplitude of 12 V on the nanopositioner. As the tip is gradually retracted the contact resistance increases, and these additional dips in the spectra gradually diminish and eventually disappear for contact resistance of 16.15 $\Omega$. Here we realise a pure ballistic contact. The fine positional control and stability of our nanopositioner is apparent from the set of curves for which the resistance of point contact has been increased in small steps from 9.56 $\Omega$ to 16.15 $\Omega$. Fig. 8(a) shows the normalised PCAR spectra for one such ballistic contact as function of temperature. We fit these spectra with Blonder-Tinkham-Klapwijk (BTK) model[44], using $\Delta$, the interface barrier parameter Z and a small broadening parameter $\Gamma$[45] (see Appendix 2) which accounts for non-thermal sources of broadening. The temperature dependence of the $\Delta$ for Nb extracted from these fits closely follows the expected BCS (Fig. 8(b)), as expected for a conventional superconductor.

**IV. Conclusion**

In conclusion, we have successfully designed and fabricated a compact inertial nanopositioner along with piezo driving control electronics. We have integrated the Z-nanopositioner into a fully automated needle-anvil type PCAR apparatus and demonstrated its operation down to 2 K. The precision and efficiency of the nanopositioner have been effectively demonstrated, proving its capability for high-performance operation. The versatile design and small form factor allow seamless integration of these nanopositioners into low-temperature scanning probe microscopes and other cryogenic applications. These nanopositioners are thus well-suited for a broad range of advanced research and industrial applications.

**Appendix 1.**

The Adafruit Feather M4 Express microcontroller receives a 9-letter string from a LabVIEW interface, which it uses to perform a particular task with the desired amplitude and frequency. First letter of the string gives the command for generating different waveforms: 'U' stands for a continuous increasing sawtooth waveform (for ramp-forward); 'D' means a continuous decreasing waveform (for ramp-reverse). The 'V' (and 'E') moves the nanopositioner by a single step in forward (reverse) direction. The 'S' command stops generating any waveform. The subsequent five characters represent the amplitude specified by the user, which corresponds to a desired output voltage. The letter 'F' follows, acting as a separator between the amplitude and the next command. The 7th to 9th characters provides the frequency settings input by the user (only used continuous forward or reverse ramps), and the final 'F' marks the end of the command. This string of commands is illustrated in Figure 9 for clarity.

## Appendix 2.

For fitting the normalized experimental spectra, we have used Blonder-Tinkham-Klapwijk (BTK) model[44]. Within this model the current across the interface is written in terms of the Andreev reflection coefficient, $A(E, \Delta, Z, \Gamma)$ and normal reflection coefficient, $B(E, \Delta, Z, \Gamma)$ of the electron from the normal metal to the superconducting side as,

$$I(V) \propto \int_{-\infty}^{\infty} [f(E - eV) - f(E)][1 + A(E, \Delta, Z, \Gamma) - B(E, \Delta, Z, \Gamma)]dE,$$

where $E$ is the energy of the incoming quasiparticle with respect to the Fermi Energy, $f(E)$ is the Fermi function. $A(E, \Delta, Z, \Gamma)$ and $B(E, \Delta, Z, \Gamma)$, calculated by solving the Bogoliubov–de Gennes (BdG) equations for a nonmagnetic metal/ superconductor interface. This model has three parameters: The superconducting energy gap $\Delta$, the interface barrier parameter Z and a small broadening parameter $\Gamma$[45] which accounts for non-thermal sources of broadening. To model a realistic interface, a delta-function potential in the form of $V_0\delta(x)$ is assumed at the interface. The delta-function approximation simplifies the mathematical treatment by concentrating the potential into a single point, capturing the essential physics of the boundary or interface without complicating the equations with more spatially extended potentials. This delta-function potential is parametrized within the model as a dimensionless parameter, $Z = V_0/\hbar v_F$ ($\hbar$ is the Plank's constant and $v_F$ is the average Fermi velocity of tip and sample). In addition to the effect of any oxide barrier that may be present at the interface, $Z$ also accounts for an effective barrier arising from the fermi velocity mismatch between the normal metal and the superconductor. The broadening parameter phenomenologically accounts for the broadening of the superconducting density of states (DOS) from its ideal BCS value. Originally introduced as a phenomenological parameter to account for a finite lifetime of the superconducting quasiparticles[46], in recent years this parameter has found wider interpretations. It has been shown that it also mimics the effect arising from a distribution of superconducting gap values[47] as well as the broadening arising from disorder scattering[48]. The full expressions for $A(E, \Delta, Z, \Gamma)$ and $B(E, \Delta, Z, \Gamma)$ can be found in ref. [49]. The resulting fit of the spectra at various temperatures is shown through the black solid lines in Fig. 8(a). The extracted information about BCS gap $\Delta$, $Z$ and $\Gamma$ from the fits are plotted as a function of temperature. As seen in Fig. 8(b), $\Delta$ follows BCS temperature dependence and $Z$, $\Gamma$ is almost independent of temperature.

***Acknowledgement***: We thank Vikas Kale for making 3D drawing of our Nanopositioners, point contact head and insert, Vivas C. Bagwe for technical help and Surajit Dutta for his involvement in the early stages of this work. We also thank the Central Workshop of the Tata Institute of Fundamental Research for machining different parts of the Nanopositioner and the Point Contact spectroscopy apparatus. We would like to thank Department of Atomic Energy, Govt of India for financial support. KKS would like to thank Homi Bhabha National Centre for Science Education for hosting her visit in TIFR under the NIUS program. SolidWorks drawing of all machined parts are available on request.

PR and JJ designed and fabricated the nanopositioners. PR, JJ and KKS developed the control electronics. PD, SD and KKS characterised the nanopositioners. PD and SD fabricated the point contact spectroscopy insert and performed PCAR measurements. PD, SD and PR wrote the paper with inputs from all authors. PR conceived the idea and supervised the project.

Table 1. Calibration parameters of Z and X nanopositioner with and without load.

| Nanopositioner | Voltage for Piezo Vp (V) | Step-size (nm) Up/Forward | Step-size (nm) Down/Backward |
|---|---|---|---|
| Z- Nanopositioner with no Load | 10 V | 80.23 | 177.86 |
| | 34V | 100.87 | 1060 |
| Z- Nanopositioner with 16.9g Load | 10 V | 135.3 | 341.81 |
| | 34 V | 788.57 | 1980 |
| Z- Nanopositioner with 21.5g Load | 10 V | 127.87 | 393.54 |
| Z- Nanopositioner with 30.9g Load | 10 V | 94.37 | 510.71 |
| | 34 V | 508.33 | 2540 |
| | 61 V | 1020 | 5170 |
| X- Nanopositioner without Load | 10 V | 225.5 | 225.5 |
| | 34 V | 981 | 967 |
| X- Nanopositioner with 4.6g Load | 10V | 177.3 | 175.6 |
| | 34V | 858 | 858 |

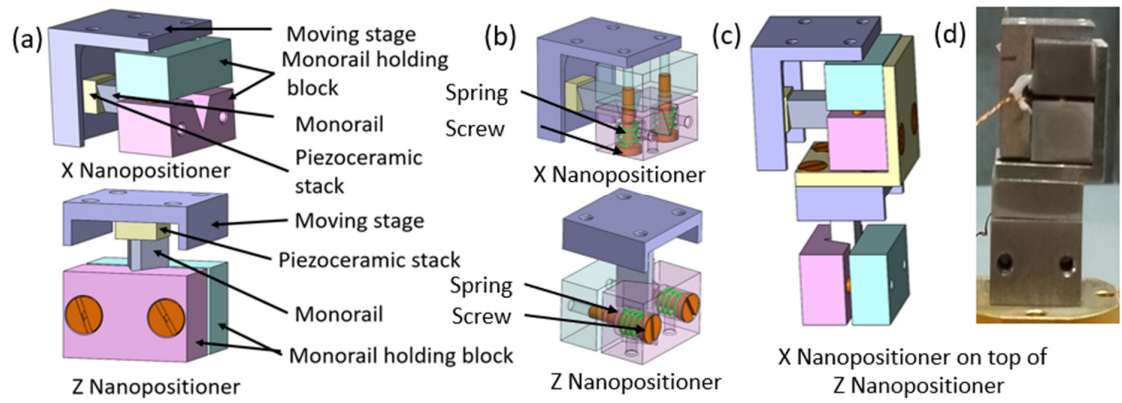

**Fig 1.** (a) Design of the X (top) and Z (bottom) nano-positioners. (b) Same as (a) except that the block holding the monorail have been made semi-transparent to show the spring arrangement. (c) Schematic of 2-axis configuration, where the X positioner is stacked above the Z positioner. (d) Photo of 2-axis X-Z nanopositioner.

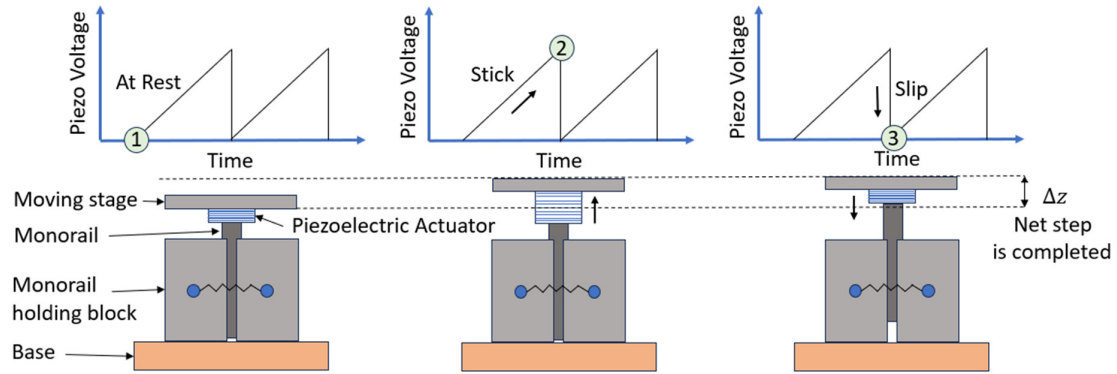

**Fig 2.** Working principle of the slip-stick positioner: (left) The piezo stack in contracted position corresponding to 1 in the upper panel. (middle) As the voltage is slowly increased the piezo-stack expands raising the moving stage up; 2 corresponds to fully expanded position. (right) When the voltage is rapidly decreased from 2 to 3, the piezo-stack rapidly contracts causing rapid acceleration of the moving stage; due to the reaction force the, triangular monorail slips, thus raising the moving stage by a finite amount Δz as shown in the figure.

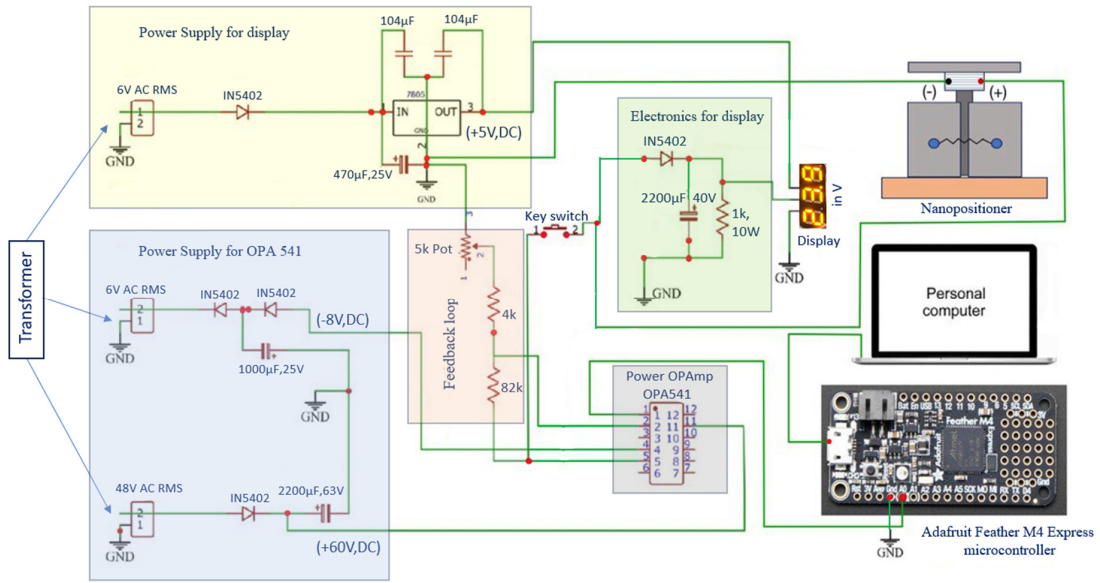

**Fig 3.** Schematic of circuit of control unit for inertial nanopositioner. The 4 main portions of the circuit are highlighted in different colors. An additional reset mechanism for faster response of the display when the amplitude of the sawtooth is changed is not shown in the diagram. When the Key switch is off the output of the circuit is isolated from the nanopositioner.

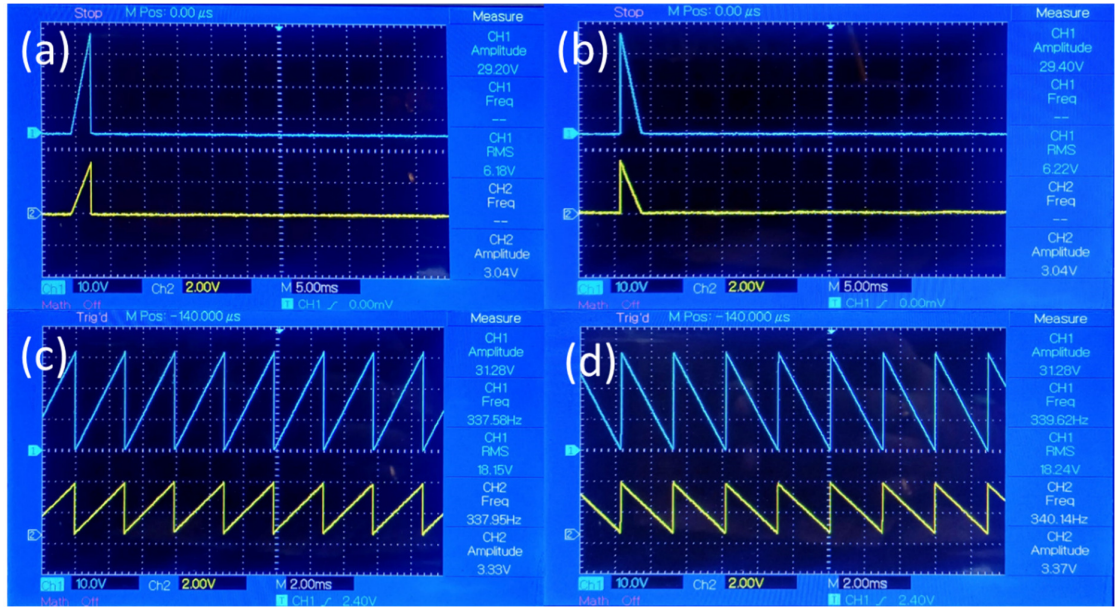

**Fig 4.** Oscilloscope traces of the waveform for moving the nano-positioner. (a) Waveform for one single step up. (b) Waveform for one single step down. (c) Waveform for continuous steps up. (d) Waveform for continuous steps down. As can be seen from the snapshots, Channel 1 Amplitude (shown in blue) with 10 V per division scale along y-axis, which is measured at the output of the piezo actuator power supply being fed to the positioner is amplified as compared to the Channel 2 Amplitude (shown in yellow) with 2V per division scale along y-axis, which is measured at the output of the Adafruit Feather M4 Express microcontroller. The sharp edge of the sawtooth is <10 microseconds.

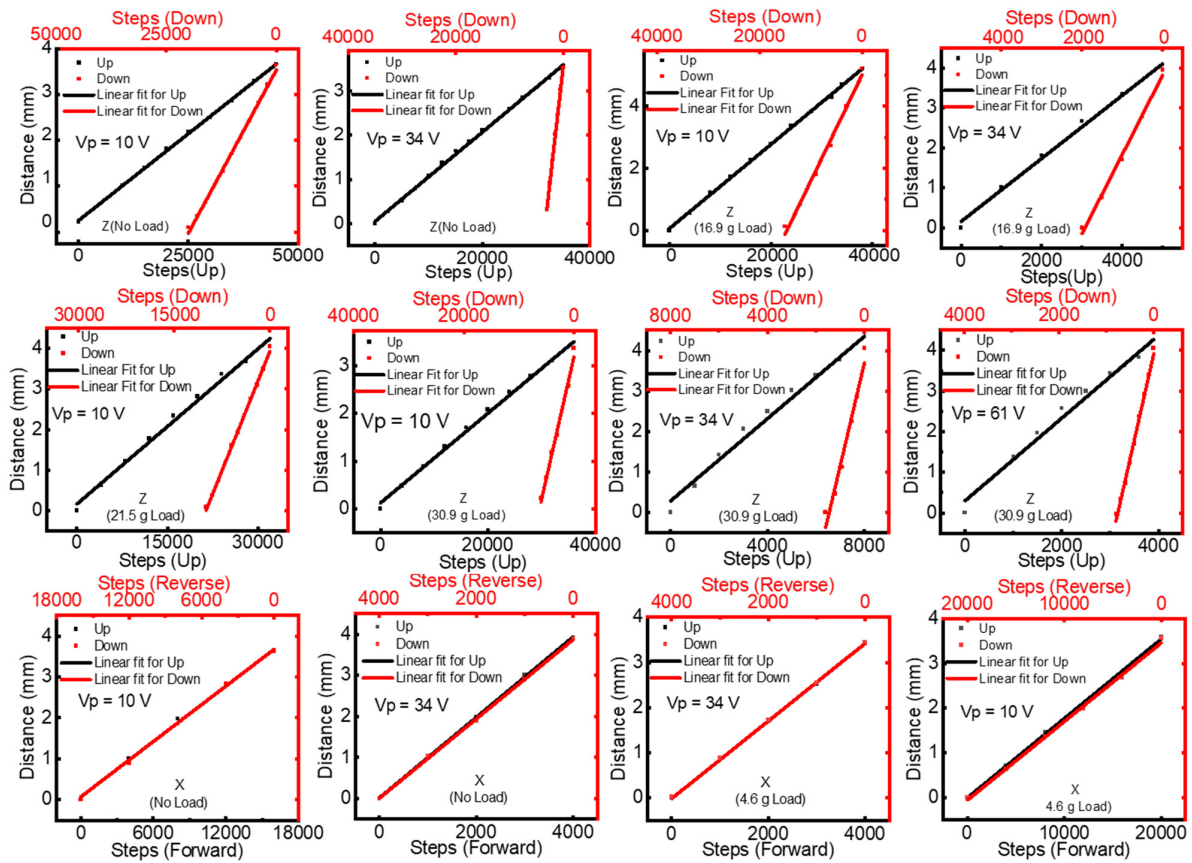

**Fig 5.** Movement as a function of steps for the Z nanopositioner with no load on it at (a) 10 V, (b) 34 V, with 16.9gram load on it at (c) 10 V, (d) 34 V, with 21.5gram load at (e) 10V and with 30.9 gram load at (f) 10V, (g) 34V, (h) 61V. same for the X nanopositioner with no load on it at (i) 10 V, (j) 34 V, with 4.6gram load on it at (k) 10 V, (l) 34 V.

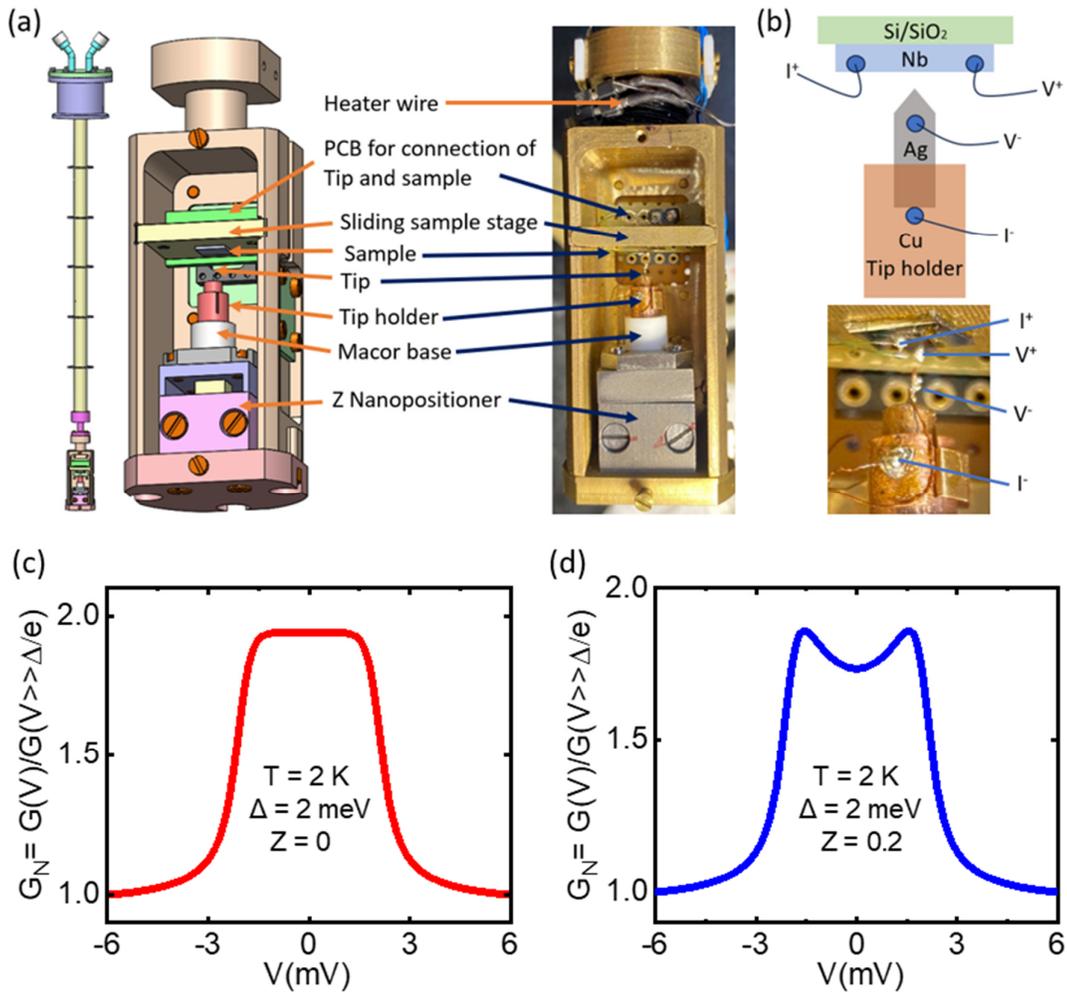

**Fig 6.** (a) Design of probe used for Point Contact Spectroscopy. (b) Schematics of electrical connections point-contact measurements. (c) Representative PCAR spectra for an ideal transparent N/S contact. (d) Representative PCAR spectra for a realistic contact with finite transparency. Parameters used to simulate the spectra are shown in the panels.

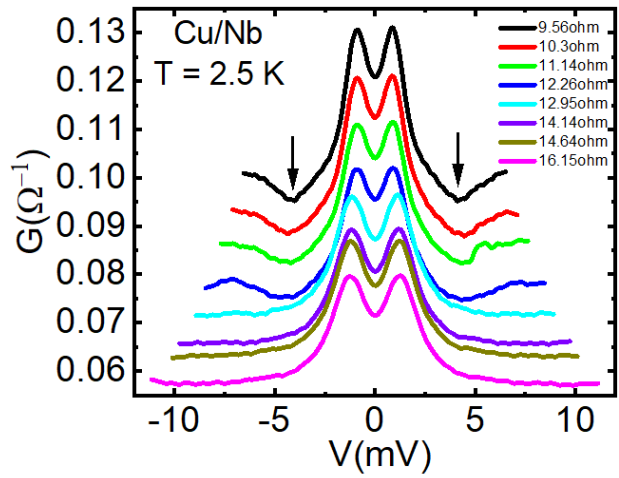

**Fig 7.** *G(V) vs V* spectra at *T* = 2.5 K for Cu tip-Nb film point contact with different contact resistance. The uppermost black curve corresponds to the spectrum right after the tip is engaged on the sample. Each successive curve below is recorded after reducing the contact diameter, by lowering the tip from the sample in small steps.

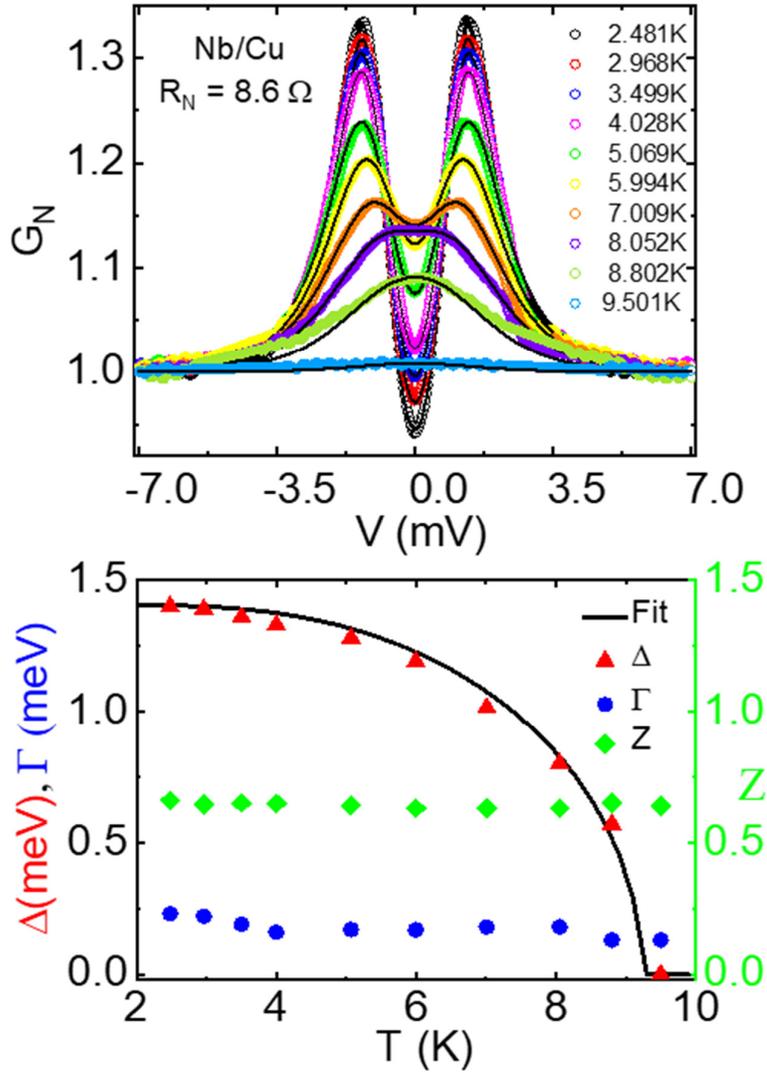

**Fig 8.** (a) Normalised conductance spectra ( $G_N$ vs $V$ ) for a ballistic contact between Cu tip-Nb film at different temperatures. The solid lines are fit to the BTK model. (b) Temperature variation of $\Delta$, $Z$ and $\Gamma$ extracted from the fits of the spectra in panel (a). The solid black line shows the theoretical variation of $\Delta$ from BCS theory.

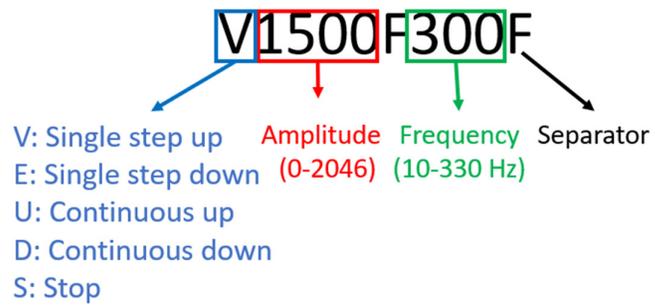

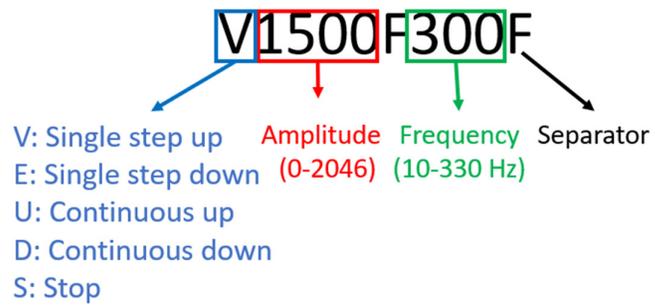

**Fig 9.** The string input going to the Adafruit Feather M4 Express microcontroller through the serial port.

# Supplementary Information

## Development of low-cost compact inertial nano-positioner operating at cryogenic temperatures

Pritam Das, Sulagna Dutta, Krishna K.S., John Jesudasan and Pratap Raychaudhuri

*Tata Institute of Fundamental Research, Homi Bhabha Road, Mumbai 400005.*

## I.    Arduino IDE Sketch:

```
const int out_pin = A0; // Define the output pin as A0
int i = 0;
int ii = 0;
int iii = 0;
int j = 0;
int amp = 4095; // Maximum value for 12-bit resolution
double f = 200.0; // Default frequency
double df; // Variable for delay time
int b = (amp + 1);
int ftest;
double freq = 0;
char ser[10];  // Stores concatenated string
char sernum[4];  // Stores amplitude value from input string
char serf[4];  // Stores frequency value from input string

void setup() {
    Serial.begin(9600); // Initialize serial communication at 9600 baud
    analogWriteResolution(12); // Set analog write resolution to 12 bits (0 - 4095)
    analogReadResolution(12); // Set analog read resolution to 12 bits (0 - 4095)
}

void loop() {
    // Check if serial input starts with 'U' for continuous up waveform
    if (ser[0] == 'U') {
        for(i = 0; i < b; i++) {
            analogWrite(out_pin, 2 * i);
            delayMicroseconds(df); // Delay
        }
    }
    // Check if serial input starts with 'D' for continuous down waveform
    else if (ser[0] == 'D') {
        for(i = 0; i < b; i++) {
            analogWrite(out_pin, 2 * (b - i));
            delayMicroseconds(df); // Delay
        }
    }
    // Check if serial input starts with 'V' for single step up
```

```
        else if (ser[0] == 'V') {
        for(i = 0; i < b; i++) {
            analogWrite(out_pin, 2 * i);
            delayMicroseconds(df); // Delay
        }
        ser[0] = 'S';  // Command to stop
    }
    // Check if serial input starts with 'E' for single step down
    else if (ser[0] == 'E') {
        for(i = 0; i < b; i++) {
            analogWrite(out_pin, 2 * (b - i));
            delayMicroseconds(df); // Delay
        }
        ser[0] = 'S';  // Command to stop
    }
    // If no valid command, turn off the output
    else {
        analogWrite(out_pin, 0);
    }

    j = 0;
    while (Serial.available() > 0) {
        ser[j] = Serial.read(); // Read the input string
        j++;

        if (Serial.available() == 0) {
            // Extract amplitude and frequency values
            for (ii = 0; ii <= 3; ii++) {
                sernum[ii] = ser[ii + 1];  // Amplitude
                serf[ii] = ser[ii + 6];   // Frequency
            }

            b = atoi(sernum); // Convert to integer
            f = atoi(serf);
            df = (1000000 / (f * b)) - 1.363; // Calculate delay based on frequency and amplitude

            // Print the values to the serial monitor
            Serial.print(ser[0]);
            Serial.print("   ");
            Serial.print(b);
            Serial.print("   ");
            Serial.print(f);
            Serial.print("   ");
            Serial.println(df);
        }
    }
}
```

## II. Details of the LabView interface:

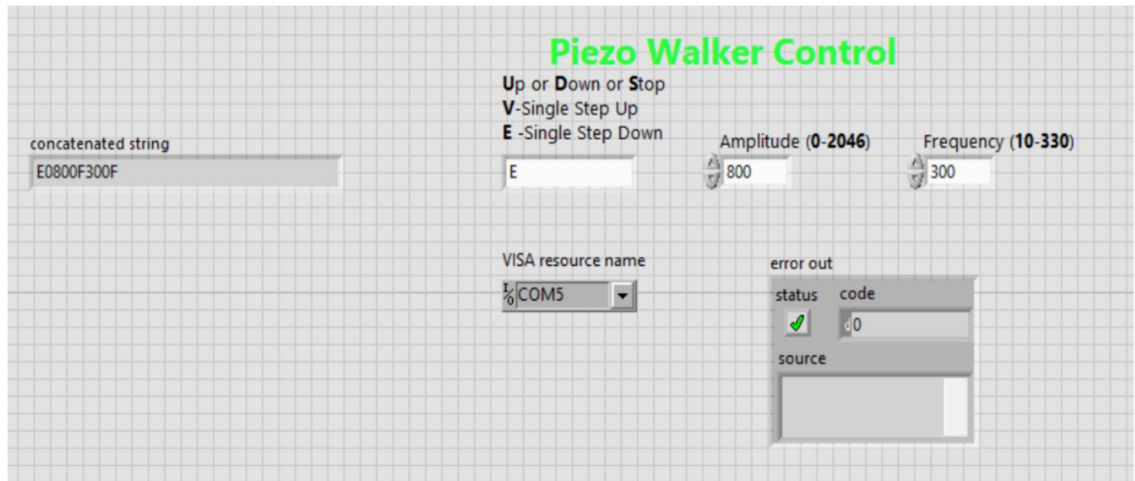

Fig S1. Front panel of the nanopositioner control sub-vi

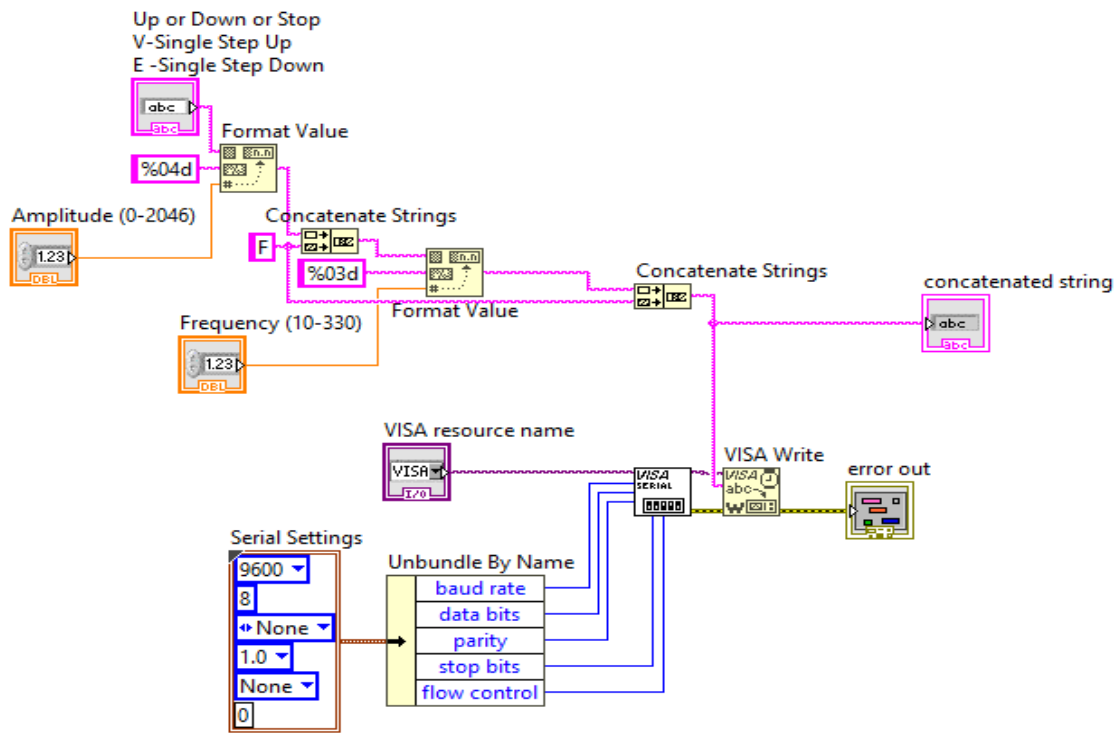

Fig S2. Block diagram of the LabView program